\documentclass[12pt,preprint,natbib209]{aastex}

\newcommand{\civ}{C~{\sc IV}}
\newcommand{\niii}{N~{\sc III}}
\newcommand{\brg}{Br$\gamma$}
\newcommand{\hei}{He~{\sc I}}

\newcommand{\kms}{km~s$^{-1}$}

\slugcomment{}

\shorttitle{Massive binaries in UC~H\sc{II} regions}
\shortauthors{Apai et al.}

\begin{document}


\title{Massive binaries in high-mass star-forming regions: 
A multi--epoch radial velocity survey of embedded O-stars\altaffilmark{1}}


\author{D\'aniel Apai\altaffilmark{2,3}}
\affil{Steward Observatory, University of Arizona, 933 N. Cherry Avenue, Tucson, AZ 85721, USA}
\email{apai@as.arizona.edu}

\author{Arjan Bik\altaffilmark{4}}
\affil{European Southern Observatory, Karl-Schwarzschild-Str. 2, D-85748 Garching, Germany}

\author{Lex Kaper}
\affil{Astronomical Institute "Anton Pannekoek", Kruislaan 403, 1098 SJ
Amsterdam, The Netherlands}

\author{Thomas Henning}
\affil{Max Planck Institute for Astronomy, K\"onigstuhl 17,  D--69117 Heidelberg,
Germany}

\and

\author{Hans Zinnecker}
\affil{Astrophysikalisches Institut Potsdam, An der Sternwarte 16, 14482 Potsdam,
Germany}

\altaffiltext{1}{Based on observations collected at the European Southern 
   Observatory at Paranal, Chile (ESO programs 64.H-0425, 65.H-0602
   and 69.C-0189)}
\altaffiltext{2}{Max Planck Institute for Astronomy, K\"onigstuhl 17, D--69117 Heidelberg,
Germany}
\altaffiltext{3}{NASA Astrobiology Institute}
\altaffiltext{4}{Astronomical Institute "Anton Pannekoek", Kruislaan
  403, 1098 SJ Amsterdam, The Netherlands}


\begin{abstract}
We present the first multi-epoch radial velocity study of 
embedded young massive stars using near--infrared spectra obtained with ISAAC mounted at the
ESO Very Large Telescope, with the aim to detect massive binaries.
Our 16 targets are  located in high--mass star--forming regions and many of them are associated to known ultracompact H{\small II}
regions, whose young age ensures that dynamic evolution of the clusters did not
influence the intrinsic binarity rate.
We identify two stars with about 90 km/s
velocity differences between two epochs proving the presence of
close massive binaries.  The fact that 2 out of the 16 observed stars are binary systems suggest that
{\em at least} 20\% of the young massive stars are formed in close multiple systems, but may also be consistent
with most if not all young massive stars being binaries.
In addition, we show that the radial velocity
dispersion of the full sample is about 35 km/s, significantly larger
than our estimated uncertainty (25 km/s). 
 This finding is consistent with similar measurements of the young massive cluster 30 Dor which might have a high intrinsic binary rate.
Furthermore, we argue that virial cluster masses derived from the radial velocity
dispersion of young massive stars may intrinsically overestimate the cluster mass due to
the presence of binaries.

\end{abstract}

\keywords{ binaries: close -- binaries: spectroscopic -- stars: early-type -- stars: formation -- infrared: stars }

\section{Introduction}
The formation and early evolution of young massive stars is not yet understood, motivating substantial observational and theoretical work.  Observationally it is clear that most, or even all, massive stars form in clusters and OB--associations \citep{2005A&A...437..247D}. After their formation the massive stars remain hidden in their natal molecular cloud. This makes the youngest phases of massive stars difficult to study as only the infrared and the X--ray windows are accessible to detect the recently formed massive stars.

Only after a few million years, the massive stars become optically visible. By this time, many of them are found in multiple systems \citep{1982ApJ...263..777G,2003Lanzarotte}.  In the Orion Trapezium cluster all but one of the massive stars are binaries with at least 1.5 companion per primary star on average  \citep{1999NewA....4..531P}.

A spectroscopic survey of the 30 Doradus cluster in the Large Magellanic Cloud suggests that most massive stars in that cluster are binaries.  The radial velocity  dispersion is much
 larger than what is estimated from the cluster dynamics  and the
 dispersion is probably entirely dominated by binary orbital
 motions \citep{2001A&A...380..137B}. Numerical simulations show that the observations are
 consistent with the hypothesis that all (or most) stars in the cluster are  binaries. 

As those clusters are already a few million years old, dynamical cluster evolution might already have severely modified the binary fraction of the clusters members. 
In order to eliminate the influence of cluster dynamics on the 
binarity rate and to observe the initial binary fraction, other  approaches have to be taken \citep{2003Lanzarotte}.

Binarity studies of massive members of OB associations  \citep[e.g.][]{2001AN....322...43B,2006ApJ...639.1069H} can only provide a biased picture on the intrinsic binarity rate due to the evolution of massive stars over the typical lifetime ($\sim$~5--20~Myr) of the associations.

For the determination of the binary rate of the most massive stars,
the study of very young ($<$1 Myr) star--forming regions is necessary.  Unfortunately, the multiplicity study of such young stars faces four serious observational difficulties: First,  the large distances toward massive star--forming regions make direct imaging of close binaries very challenging and often impossible.  Second, the
youngest massive stars are deeply embedded and are only visible in the
infrared regime. Third, these stars have very few and weak spectral lines in the infrared; in fact, they are often used as spectroscopic flat field
sources. Fourth, being rapid rotators their spectral lines are broad
and occasionally blending.

\begin{figure}
  \centering
\plotone{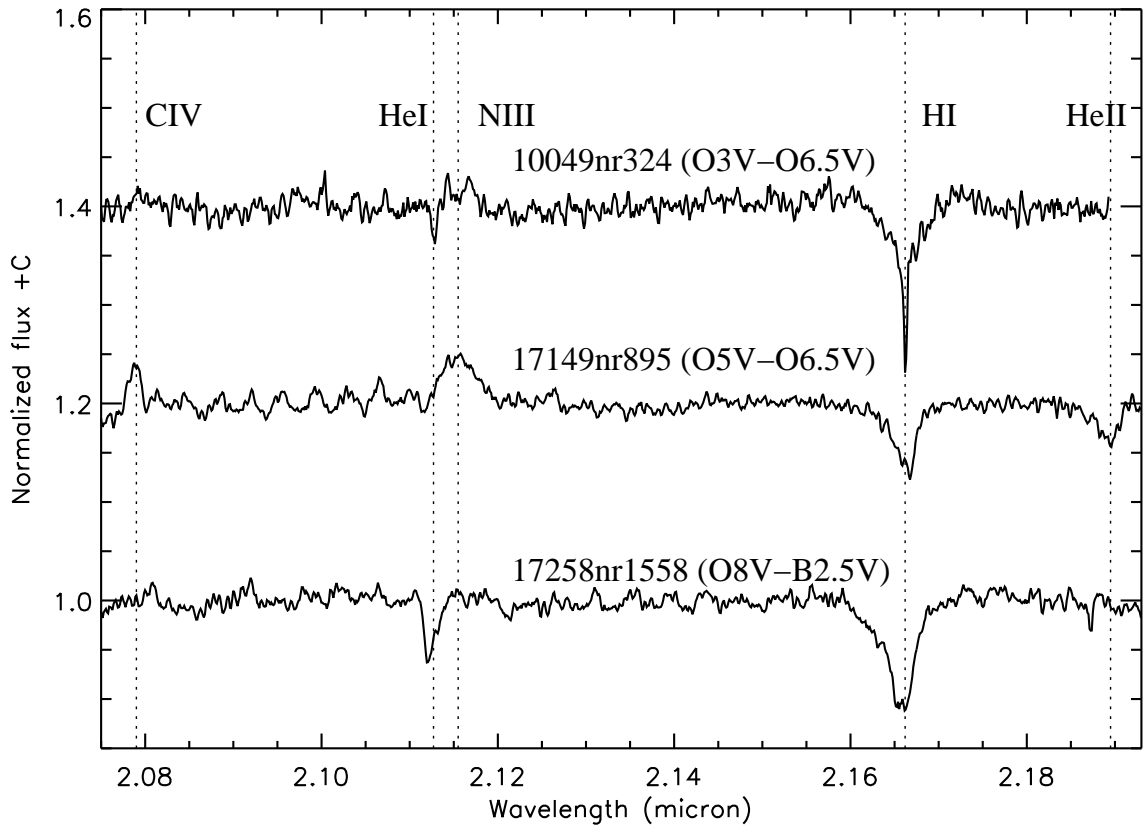}
     \caption{Representative VLT/ISAAC K--band spectra of OB stars: the CIV line is
     at 2.078 $\mu$m, the He I line is at 2.113, the NIII line is at 2.116 $\mu$m, the HI is at 2.166, the
     HeII is at 2.189 $\mu$m.}
	\label{RVSpecExam}
  \end{figure}

The recent years, however, proved that the line system in the K--band
spectrum of massive stars, when observed with high enough signal--to--noise ratio and
spectral resolution, still provides reliable spectral type
classification \citep{1996ApJS..107..281H,2005ApJS..161..154H}.  The K--band spectral atlas
has been applied by \cite{2005A&A...440..121B} to 40 high-mass stars in high--mass star-forming regions, mostly close to ultracompact H~{\small II} regions using the ISAAC/VLT
infrared instrument.  To demonstrate the K--band spectra of young OB stars we show  some representative spectra from this survey in Fig.~\ref{RVSpecExam}.  The \cite{2005A&A...440..121B}   survey yields a
unique spectroscopic data set on the youngest observable massive
stars.

Aiming to determine the intrinsic multiplicity of the massive stars we
obtained second-- and occasionally third--epoch spectra for 16 targets of
this sample using an identical instrument setup. By identifying the
variability of the massive stars' radial velocities we probed the
presence of close, massive companions and identified at least two embedded, close, and massive 
multiple systems. 

\section{Observations and data reduction}
\label{Observations}

Our target list consists  of spectroscopically confirmed massive stars
identified via near--infrared photometry. \citet{Kaper06} carried out 
multi--color NTT/SOFI imaging of 44 IRAS sources with colors typical
to ultracompact H{\small II} regions and identified young massive star candidates
via their position on the color-magnitude diagram (red and bright). 
In 31 of these regions \citet{2005A&A...440..121B} carried out near--infrared spectroscopy
confirming 38 massive stars.  

This survey provided spectral type classification of the massive stars and served as the first-epoch
radial velocity measurement for our survey. By re-observing the most massive stars (O and early B) we
extended these measurements into a multi--epoch radial velocity
survey. In order to maximize the homogeneity of the data at different
epochs we used the same slit positions for all but three objects. In
these three cases only one massive star was covered in the original
slit position: the new slit orientation allowed probing additional
bright stars in the regions. For five stars we were able to obtain an 
additional, third--epoch measurement.

 Following the nomenclature adopted in
\citet{Kaper06} and \citet{2005A&A...440..121B} we
will name the objects after the first 5 digits (i.e. the right
ascension) of the IRAS point source they are associated with,
followed by a running number from our initial photometric identification  (e.g. object 324 in
IRAS 10049-5657 we refer to as 10049nr324).

\subsection{ISAAC Observations}

The spectroscopic data presented in this paper were obtained during
several runs between March 2000 and June 2002 using the
1024$\times$1024 pixel ISAAC near--infrared camera and spectrograph
\citep{1997SPIE.2871.1146M} mounted at the Nasmyth focus of the UT1 of the
Very Large Telescope of the European Southern Observatory at Cerro
Paranal, Chile.  Table~\ref{T/Observations} lists the targets and
the journal of observations

\begin{table*}
\begin{center}
\caption{Journal of observations and the target list. \label{T/Observations}}
\begin{tabular}{llccc}
\tableline\tableline
 IRAS Source  &  Slit ID &  1st Epoch & 2nd Epoch & 3rd Epoch  \\
\tableline
\object{IRAS 10049-5657} & Single  &  00/03/19 & 02/06/19	 & 2 $\times$ 02/06/20   \\
\object{IRAS 15408-5356} & Single  &  00/03/20 & 02/06/19	 & 02/06/20              \\
\object{IRAS 16177-5018} & Slit 1  &  00/05/08 & 02/06/19	 & 02/06/20              \\
\object{IRAS 16177-5018} & Slit 2  &  00/05/08 & 02/06/19	 & 02/06/20              \\
\object{IRAS 16571-4029} & Single  &  00/05/08 & 02/06/20 	 &---                    \\
\object{IRAS 17149-3916} & Single  &  00/05/19 & 02/06/19	 &---                    \\
\object{IRAS 17258-3637} & Single  &  00/05/19 & 02/06/19	 &---                    \\
\object{IRAS 18449-0158} & Slit 1  &  00/06/20 & 02/06/19	 & 02/06/20              \\
\object{IRAS 18507+0110} & Slit 1  &  00/07/12 & 02/06/19	 &---                    \\
\object{IRAS 18507+0110} & Slit 2  &  00/06/20 & 02/06/19	 &---                    \\
\tableline
\end{tabular}

\end{center}
\end{table*}

The observations were carried out in a standardized fashion to ensure
the homogeneity of the data set.  For all objects the observed
spectral range extends from 2.075~$\mu$m to 2.2~$\mu$m, centered on
2.134~$\mu$m with three exceptions: During the first epoch
observations the objects IRAS 10049-5657 and IRAS 15408-5356 have been
observed with a spectral range centered at 2.129~$\mu$m, while in the
second epoch the object 10049-5657 has been observed additionally with
a spectral range centered at 2.138~$\mu$m to also cover the He II line at 2.185~$\mu$m.  In order to reach the
highest spectral resolution we used the smallest available slit width
of 0.3\arcsec. The effective resolution with a resolving power of R
$\simeq$ 10,000 corresponds to a radial velocity resolution of
$\simeq$ 30 km/s.

In order to free the spectra from Earth's atmospheric influence, after
each science target, we observed a telluric standard with an identical airmass.  As telluric standards we used A--type stars, which display no
lines apart from the Br$\gamma$ line in the spectral regime in
question. After every standard star observation a flat field has been
obtained in order to correct for fringing patterns which become
visible in high S/N K-band spectra.

\subsection{Data Reduction}\label{DataReduction}

The spectra from the first and the second epoch are reduced in an identical way in order to obtain  more homogeneous measurements. A detailed description of the reduction process  is given in   \citet{2005A&A...440..121B} and in \citet{2004PhDT.........1A} and is performed using standard IRAF and
IDL routines. In the following we will only briefly
overview the non-standard and critical reduction steps, such as the
wavelength calibration and the telluric line removal.

The pixel--to--wavelength calibration is performed using the OH emission line spectrum present in the observed spectra \citep{2000A&A...354.1134R} with the IRAF task \emph{identify}.
The derived wavelength solution was typically better than 0.2 pixel (3 \kms). This wavelength solution was then
step-by-step shifted and matched in the spatial direction along the
slit. An independent wavelength calibration has been derived for each night.

The telluric standard stars, observed at matching airmasses (with typical differences less than 0.1), are 
used to remove the telluric absorption lines in the spectra. We applied main sequence A--type 
stars as telluric standards because their K--band spectra are featureless except for the Br$\gamma$ line.
The \brg\ line was removed in two steps:  First, the telluric features are removed from
the $K$--band spectrum of the telluric standard using a high--resolution
telluric spectrum (obtained at NSO/Kitt Peak). This spectrum is taken
under very different sky conditions, so numerous spectral remnants remain
visible in the corrected standard star spectrum. Without this
``first-order'' telluric--line correction, however, a proper fit of \brg\ cannot
be obtained. In the second step the \brg\ line was fitted by a combination of two
exponential functions. The error on the resulting \brg\ equivalent
width  of our target star was about 5~\%. After the removal of the
\brg\ line of the A star, the telluric lines are removed using the IRAF task \emph{telluric}. 
The task applies a cross-correlation procedure to determine the
optimal shift in wavelength and the scaling factor in line strength,
which could be adjusted interactively. The shifts were usually a few
tenths of a pixel; also the scaling factors are modest ($\sim 10$~\%).

\section{Radial Velocity Measurements}
\label{S:RVmeasurement}
  The final step of the procedure was the measurement of the radial
 velocity of each object at a given epoch. As our
 science goal is the detection of close binaries, we were focusing on
 deriving relative radial velocities and did not aim at deriving the
 absolute values. For the sake of simplicity, we always chose the
 first epoch observation (see, Tab.~\ref{T/Observations}) as the radial
 velocity zero point and compared the subsequent epoch(s) to this
 reference value by using the IRAF package {\it RV} and the {\it
 fxcor} task. 
 
This package is widely applied for accurately
 measuring the radial velocities by means of cross-correlation. As the
 package has been discussed by \citet{1994adass...3...79F} and
 \citet{1998Ap&SS.263..259R}, we only summarize here a few essential
 points and non-standard choices of parameters.
 
 The radial velocity is determined by cross-correlating the first
 epoch measurement to the later epoch ones, after correcting for any differences in 
heliocentric velocities. However, only the
 photospheric lines of the massive stars carry valuable
 information on the radial velocity. The continuum is usually
 dominated by telluric line remnants and might influence
 the cross-correlation.  Therefore, the individual spectral regions of interest
 have been selected manually after the inspection of each reduced
 spectrum. In practice, the \hei\ and \brg\ lines are mostly used for
 the cross-correlation (see, Appendix \ref{objects} for details per object).
   
\begin{figure}
   \centering
   \plotone{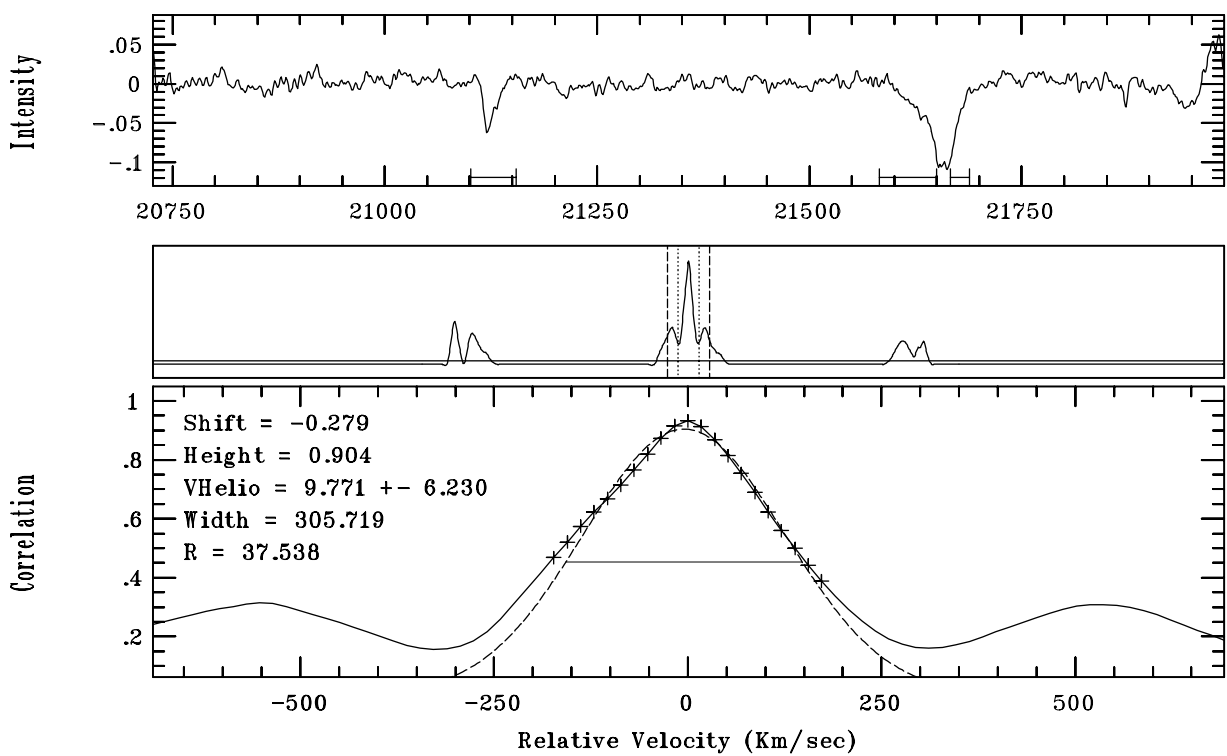}
   \caption{Example for the cross--correlation based relative radial velocity
   determination. The upper panel shows the object spectrum and the horizontal
   bars mark the wavelength range which was used for the cross-correlation. 
   The middle panel shows the complete cross-correlation function, while
   the lower one displays a magnified region around the peak. The dashed curve
   is the fitted Gaussian and the horizontal solid line is its full width at
   half maximum. This plot is the output of the IRAF script {\it fxcor}.
   \label{F:CorrelationExample}}
      \end{figure}

 The actual radial velocity difference ($\Delta$RV), i.e. the wavelength shift
 between the two spectra is derived from the cross-correlation
 function: Its maximum is where the shifted second epoch spectrum
 matches best the first epoch one.  The maximum of the
 correlation function was identified by fitting a Gaussian function on
 the peak. To adopt our measurement to the rather broad lines of the
 massive stars we enlarged the default short fitting interval to
 a maximum of 100 pixel interval.  Fig.~\ref{F:CorrelationExample} shows
 an example for the radial velocity determination and the fitting of
 the cross-correlation function.
 
\subsection{Error Analysis}
\label{S:Errors}
The observations described in this paper represent a pilot effort to
derive radial velocity variations for deeply embedded massive
stars. These measurements are complex making a detailed error analysis
essential.  In the following we discuss the impact of the individual
error sources and measure the total error from quasi-simultaneous
observations.

\subsubsection{Telluric Line Residuals}
\label{S:TelluricResiduals}

 All our spectra bear to some extent the imprint of Earth's atmosphere in
the form of fixed--wavelength telluric lines. Although the individual
line residuals have modest strength, their integrated power could bias the
$\Delta$RV measurements towards the zero values.

In order to test this possibility, we repeated the radial velocity
measurements but used a photospheric--line--free, pure continuum
spectral region between 2.125 -- 2.1500~$\mu$m for the
cross-correlation. This region is dominated by low--level telluric line
residuals. If the cross--correlation would be influenced by the
presence of the residuals of the given telluric lines, the radial
velocity difference would give a systematic signal close to
$\Delta$RV=0. In contrast, the cross--correlation returns randomly
distributed values between $\sim$ 1100 and -1100 km/s indicating that
the telluric line residuals do not introduce any systematic high-power
signal to the cross--correlation function and therefore have a
negligible influence on the real, high--power correlation functions of the stellar photospheric lines.

\subsubsection{Nebular Line Contamination}

 \brg\ and \hei\ emission lines are often present in the nebula surrounding young
massive stars and may bias simple RV--determination. The analysis
of the nebular spectra close to our sources has demonstrated that these lines are 
unresolved at our resolution.
The \hei\ line's typical intensity in the nebular spectra is only $\sim$1\% of that of the line
in the stellar spectra and thus has only a negligible effect on our RV--determination.  In order
to ensure that the more intense \brg\ does not bias our RV--determination, we excluded
the  \brg\ line center from the cross--correlation of our stellar spectra thereby relying only
on the non--contaminated wings of the \brg\ line.

\subsubsection{Wavelength Calibration}

We derived a wavelength calibration for each night of the
observations. The error of the wavelength solution is well
characterized by the comparison to the OH-line catalogue containing
thousands of narrow spectral lines. The wavelength calibration was
typically better than 0.2 pixel, corresponding to a RV inaccuracy of
3.4 km/s for unresolved lines.

\subsubsection{Peak Fitting Methods and their Errors}

 The  $\Delta$RV is determined by maximizing the
 cross--correlation function of the spectra from different epochs (see
 also Sect.~\ref{S:RVmeasurement}).  Localizing the maximum is
 accomplished by fitting the peak of the cross--correlation function.
 We tested different peak--fitting methods, including Gaussian,
 parabolic, and Lorentzian fits, as well as IRAF's Center1D routine.
 All algorithms but the Center1D provide formal error estimates for
 the fit results.  In the next section we discuss the results of the different fitting methods.
 
For our data set we found that the Gaussian,
 Center1D, and Lorentzian methods gave statistically very similar
 results.  Although the
 parabolic approximation led to consistent results for most of the
 objects, in some cases it provided an unstable, out--of--bounds
 solution.
 
  Throughout the paper we used the Gaussian fitting, as this provided
 the most robust results. The fitted regime was 100 pixels
 corresponding to about 1700 km/s velocity difference ensuring that
 also the wings of the cross--correlation peak are included.

\subsubsection{Quasi--Simultaneous Observations}
\label{QSO}
For four objects we obtained subsequent spectra in two slit positions
 with typical time differences of $\sim$1~hr. Assuming that the RVs
 did not change significantly between the two observations, these 
 quasi--simultaneous observations give a good impression of the accuracy with which we are
 able to measure radial velocity differences.
 
 Using these pairs of observations we tested the consistency of the
 different RV determination methods (see, Table~\ref{T/QSOMethods}).
 In our test we used {\em direct} comparison (cross--correlation of two spectra) and
 {\em indirect} comparison (comparison of the deviations from a fiducial zero point, i.e. the
 first epoch RV). We also varied the function fitted to determine the peak of the cross--correlation
 functions in combination with the direct and indirect comparisons.

 Assuming that the radial velocity of our four stars {\em does not change}
 on timescales of an hour, all measured RV differences should be
 zero (see, Table \ref{T/QSOMethods}). The higher values seen here arise from two effects: the error
 of the RV determination and the possibility of RV variations due to
 stellar multiplicity. If the RV changes are negligible (i.e. none of the four stars are close binaries), the
 observed differences are characteristic for the measurement error;
 otherwise they overestimate the errors. Lacking the knowledge on the multiplicity of
 these four stars the RV differences derived in the following should be taken as upper limits
 for the measurement errors.

 The means of the absolute differences for the different fitting methods are listed in Table~\ref{T/QSOMethods}.  The direct method has the smallest error (19 \kms), while among
 the indirect comparisons the Gaussian method (24.2 \kms) is the most accurate and
 robust. By understanding the differences between the direct and
 indirect methods we can estimate the error budget from the
 different comparison steps: The direct comparison of two
 quasi-simultaneous spectra contains the statistical errors on 2
 measurements and the error of 1 cross-correlation and peak
 fitting. The indirect Gaussian method is carrying the errors from 3
 measurements, 1 calibration difference and 2 correlations.
 
 Our results as presented in Sect.~\ref{RVResults} use the first epoch
 spectra as reference for the later observations and therefore
 contain the errors from 2 measurements, 1 calibration difference and
 2 correlations. Their total error is in between the errors of the
 quasi-simultaneous direct and indirect Gaussian methods described
 above. Thus, the total error on our RV-difference measurements is
 between 19.0 km/s and 24.2 km/s, somewhat higher than the 14.7 km/s
 mean of the formal error estimates of the IRAF routine fxcor.

 \begin{table*}
\begin{center}
\caption{Radial Velocity differences of the four quasi-simultaneous
      observations as measured by different methods. The first column
      is the identification of the observations as given in
      Table~\ref{T/Observations}, the second column gives the epoch when
      the quasi--simultaneous observations were obtained, the  
      third column gives the RV
      difference derived from fitting a Gaussian function on the peak
      of the cross-correlation function of the two spectra. The other
      columns are the RV differences obtained by first correlating
      each second and third epoch spectra to the first epoch one and
      making the difference of the RV shifts. The last three columns
      differ in the function used for finding the peak of the
      cross-correlation function. \label{T/QSOMethods}}

\begin{tabular}{rccccc}
\tableline\tableline
 Object     & Epoch    & Direct Gaussian &  Gaussian &  Center 1D             & Lorentz  \\
            &           &   [\kms]        &  [\kms]   &  [\kms]                & [\kms]   \\
\tableline
10049nr324  &  3        & --18.0           &   --18.0   &   --11.2                &  --15.7   \\
10049nr411  &  3        &  --4.6           &    39.9   &    56.9                &   40.0   \\
16177nr1020 &  2        &  15.2           &     3.8   &   --18.8                &   --7.1   \\
18507nr373  &  2        &  38.1           &    35.1   &   ---\tablenotemark{a} &   39.4   \\
\tableline
Mean of     & Absolute Differences &     19.0           &    24.2   &    29.0                &   25.0    \\
 	 \tableline
	 \end{tabular}
	 \tablenotetext{a}{The Center1D function did not converge in this case.}
	 \end{center}
\end{table*}

\section{Results}
\label{RVResults}
The main results of our multi-epoch spectroscopic campaign are 
summarized in Table \ref{T/RVResults}. This table provides a compilation of the measured radial velocity variations, 
the spectral types  as derived by 
\citet{2005A&A...440..121B}, as well as the epochs used for the cross--correlation. We will first show the results on 2 individual objects where we have detected a large velocity difference. After that the distribution of the radial velocities is presented.

 \begin{table*}
\begin{center}
\caption{Results of the RV cross--correlation.  The columns list the IRAS source, the target ID, 
spectral type, night of the second epoch, radial velocity change relative to the first epoch
      measurements and their formal errors.
 \label{T/RVResults}}
\begin{tabular}{lllcc}
            \tableline
	    \tableline
IRAS source        & Star          &  Sp. Type       & Epochs  & $\Delta$ RV [km/s]      \\
            \tableline
IRAS 10049-5657    & 10049nr324    &   O3V -- O6.5V  & 1,3      &  14 $\pm$  11  \\
                   & 10049nr324    &   O3V -- O6.5V  & 1,3      &  32 $\pm$  12  \\
                   & 10049nr324    &   O3V -- O6.5V  & 1,2      &  15 $\pm$  14  \\
                   & 10049nr411    &   O3V -- O4V  & 1,3      &  27 $\pm$  14 \\
                   & 10049nr411    &   O4V -- O4V  & 1,3      & -13 $\pm$  29 \\
                   & 10049nr411    &   O3V -- O4V  & 1,2      &  87 $\pm$   5  \\
IRAS 15408-5356	   & 15408nr1410   &   O5V -- O6.5V  & 1,2      &  13 $\pm$  20 \\
	           & 15408nr1410   &   O5V -- O6.5V  & 1,3      &  27 $\pm$  17 \\
	           & 15408nr1454   &   O8V -- B2.5V  & 1,2      &  21 $\pm$  14 \\
                    & 15408nr1454   &   O8V -- B2.5V  & 1,3      &  18 $\pm$  25 \\
IRAS 16177-5018	   & 16177nr1020   &   O5V -- O6.5V  & 1,3      & -13 $\pm$   9    \\
	           & 16177nr1020   &   O5V -- O6.5V  & 1,2      & -19 $\pm$  14   \\
	           & 16177nr1020   &   O5V -- O6.5V  & 1,2      & -16 $\pm$  16   \\
	           & 16177nr271    &   O8V -- B2.5V  & 1,3      &  33 $\pm$  10   \\
	           & 16177nr271    &   O8V -- B2.5V  & 1,2      &  16 $\pm$   8    \\
                   & 16177nr405    &   O5V -- O6.5V  & 1,2      &  -2 $\pm$  11   \\
IRAS 16571-4029	   & 16571nr820    &   O8V -- B1V    & 1,3      &  32 $\pm$   6  \\
IRAS 17149-3916	   & 17149nr895    &   O5V -- O6.5V  & 1,2      &   1 $\pm$  10 \\
IRAS 17258-3637	   & 17258nr1558   &   O8V -- B2.5V  & 1,3      &   6 $\pm$   8 \\
                   & 17258nr378    &   O8V -- B1V    & 1,3      & -95 $\pm$  12 \\
IRAS 18006-2422	   & 18006nr770    &   O8V -- B1V    & 1,3      & -13 $\pm$  15  \\
IRAS 18449-0158	   & 18449nr319    &   O5V -- O6V  & 1,2      & -29 $\pm$  24 \\
                   & 18449nr319    &   O5V -- O6V  & 1,3      & -30 $\pm$  24 \\
IRAS 18507+0110	   & 18507nr262    &   O5V -- O8V  & 1,2      & -35 $\pm$  14	\\
                   & 18507nr373    &   B1V -- B2.5V  & 1,2      & -24 $\pm$  24 \\
         	   & 18507nr373    &   B1V -- B2.5V  & 1,2      & -59 $\pm$  18 \\
        	   & 18507nr389    &   O8V -- B2.5V  & 1,2      &  -2 $\pm$  10  \\
            \tableline							             
         \end{tabular}
	    
\end{center}
   \end{table*}

\subsection{Two compact massive binaries}

Two stars, 17258nr378 and 10049nr411, show radial velocity variations 
as large as $\sim$90 km/s between the two measurements. Variations 
of such a large amplitude can be most plausibly explained by the assumption 
that these stars are close massive binaries.  

The orbital parameters of these two systems cannot be derived from the only two epoch measurements available, but the amplitude of the RV--change suggests close, massive companions.
For example, considering the case of 17258nr378 with a primary mass of $\sim$ 15 M$_\odot$ and
assuming a circular orbit with an inclination of 70$\degr$, the
observations are consistent with a secondary of mass 9.5~M$_\odot$ and
orbital radius less than $\sim$0.4 AU.  Star 10049nr411 is a much more
massive star  (M=40--60~M$_\odot$) and the observed large $\Delta$RV requires a companion
even more massive or on a shorter--period orbit than the secondary in
17258nr378. As an example, a 50 and 20 M$_\odot$ binary at a separation
of 0.5~AU and an inclination of 70$\degr$ would be consistent with the measured radial velocity
variation.

The observations of these two sources demonstrate that at least 20\% of the 
young massive stars are in compact binary systems with massive secondaries.
We discuss the implications of this finding in Sect.~\ref{Discussion}.

\subsection{A population of young, massive binaries?}

\begin{figure}
   \centering
   \plotone{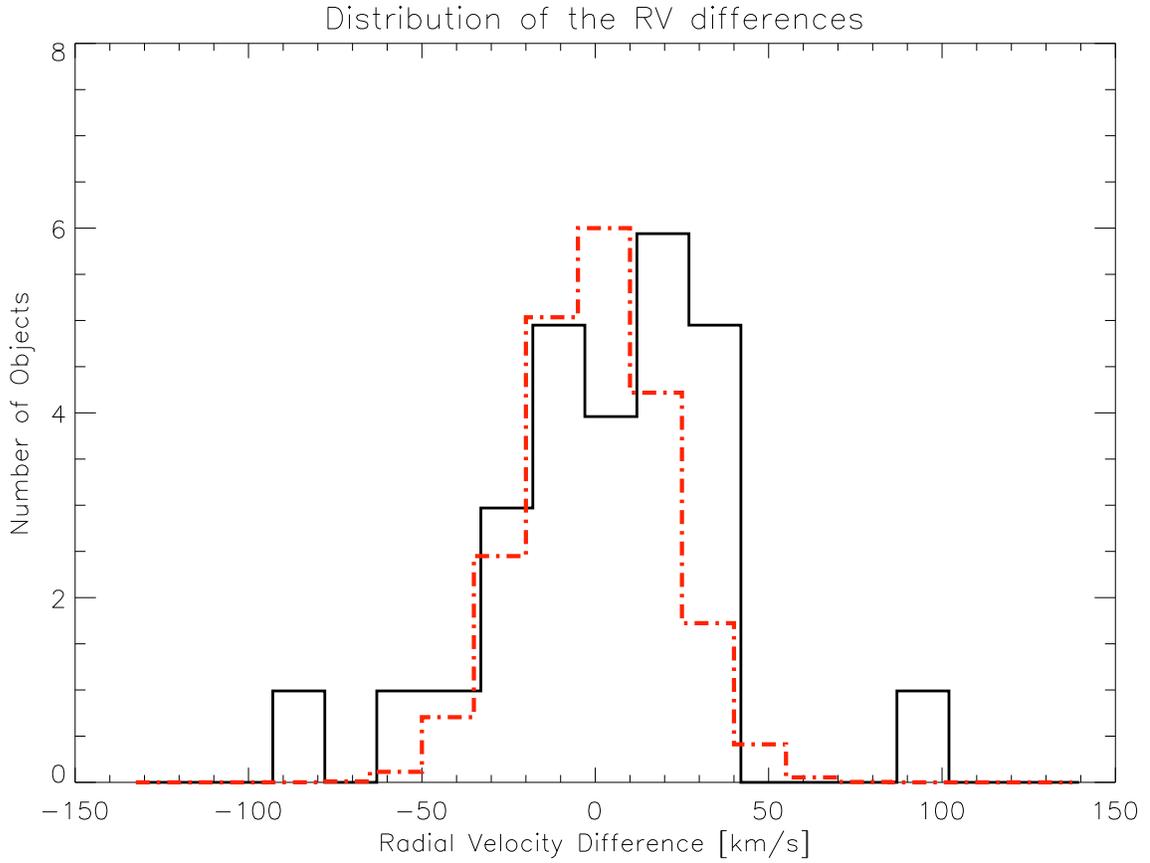}
   \caption{Histogram of the observed radial velocity differences of
   the systems at different epochs plotted with the solid black
   line. The two outliers are very good candidates for being close
   binary stars.  The dot--dashed  line indicates a hypothetic
   distribution of a single star population observed with a normally
   distributed measurement error 20 km/s. }
    \label{Hist1}
    \end{figure}

The distribution of the RV-differences is shown in the histogram in
Fig.~\ref{Hist1}. The mean value of the 27-point sample is 0~km/s 
and the standard deviation is 34.5 km/s. This is
significantly larger than the mean error which we estimated 
 to be between 19 and 24 km/s (see, Sect.~\ref{QSO}). 

If the two binary stars identified above, 17258nr378 and  10049nr411, 
are removed from the sample the standard deviation of the 
remaining measurements is 24 km/s, 
marginally consistent with our most pessimistic error estimates.

Although the RV-distribution is likely broader than the expected 
distribution of single stars, the limited number of sources that could be observed
in our survey does not allow to detect a statistically significant difference from
a single star population observed with our radial velocity accuracy.

Future infrared surveys using higher resolution instruments  -- such as VLT/CRIRES with a resolution of 100,000 corresponding to 3 km/s --- will be able to reliably discriminate between the population of single and binary stars and place constraints on the fundamental properties
of the typical compact binaries, such as mass and semi--major axis.

\section{Discussion}\label{Discussion}

In the previous section we have shown that we have identified two compact massive binary systems. Additionally, the distribution of the radial velocity measurements is larger than the error on the measurements. This might reflect the identification of a massive binary population. In this section we compare our results with other radial velocity studies and discuss the possible implications of our findings. 

\subsection{Previous binarity studies of massive stars}

Our survey probes the multiplicity of massive stars at the youngest ages and complements well
existing surveys which mainly focused on optically visible and thus somewhat older massive stars. 

The best--studied massive stars are arguably the members of the Trapezium system. 
These massive stars in the $\sim$ 1 Myr--old Orion Nebula Cluster have
at least 1.5 companions per primary star on average, as shown by
 bispectrum speckle interferometry \citep{1999NewA....4..531P}.
However, it is yet unclear, how well this binary rate represents the
intrinsic binarity of these massive stars and whether or not cluster dynamics
have already changed the structure of the Orion Nebula Cluster (see, e.g.
\citealt{1998ApJ...492..540H,1998MNRAS.295..691B}).

 An ideal target for measuring the binarity rate for a large number of
 young massive stars in an identical environment is the massive stellar
 cluster R136, which ionizes the Large Magellanic Cloud starburst
 cluster 30 Doradus.  This cluster shows three distinct peaks in its
 star formation history at 5 Myrs ago, 2.5 Myr and $<$1.5 Myr ago
 \citep{1999A&A...347..532S}. The spectroscopic survey of
 \citet{2001A&A...380..137B} presented reliable RVs for 55 stars in the 30 Dor and
 found a RV dispersion of $\sim$ 35 km/s. This dispersion is much
 larger than what is estimated from the cluster dynamics and the
 dispersion is probably entirely dominated by binary orbital
 motions. Numerical simulations show that the observations are
 consistent with the hypothesis that all stars in the cluster are
 binaries. 
 
Although this work provides a strong argument for the intrinsically
high binary rate, the 30 Dor cluster is unlikely to be representative for the
galactic massive star--forming environments due its violent starburst
nature and the multiply--peaked star formation history.  Additionally,
because of the multiply--peaked star formation process in this cluster,
the dynamically formed binaries may contaminate the intrinsic binarity
rate to an unknown extent.
 
 This latter velocity dispersion is virtually identical to the dispersion
 we find in our multi--epoch, multi--cluster survey. Although a larger sample
 is required to firmly separate the contributions from the measurement 
 uncertainties and the multiple systems, our result suggests that the 
 velocity dispersion is the same or smaller for even younger systems and
 supports  the interpretation by \citep{2001A&A...380..137B}  of close binary systems
 as the origin of the radial velocity dispersion.
 
Detecting double--line spectroscopic binaries (SB2) requires identifying the two line systems arising
from the two stellar components only possible with a spectral resolution high enough 
compared to the average separation and width of the lines.
The very few and very broad lines of massive stars  makes such detections considerably more
difficult than the detection of single--line (SB1) binaries. Due to the modest spectral 
resolution (R$\sim$10,000) and limited contrast our study is thus largely insensitive to double--line spectroscopic binaries (SB2) and 
could detect only single--line (SB1) binaries. 
However, optical spectroscopic studies of more evolved
OB--stars suggest that the frequency of SB1 binaries are similar or comparable to that of SB2 (e.g. 
\citealt{1998AJ....115..821M}). Thus, the complete binary fraction of the target sample is likely
exceeding the lower limit established through the current survey.

\subsection{The effect of massive binaries on cluster virial mass estimates}

The large variations of the RVs of the young massive stars in our sample 
demonstrate the effect of massive binaries on the cluster's RV dispersion.  We showed that
excluding the known binary systems decreases the RV dispersion from 35 km/s to 24 km/s and
possibly even more, demonstrating that in a random sample of $\sim$20 young stars at least 1/3 of the RV
dispersion is caused by binaries.  

The RV dispersion of unresolved extragalactic clusters is  a frequently used tool to derive
virial mass estimates (e.g. $M \approx \Delta RV^2$). The light of massive clusters is dominated by its most massive members.
We point out, that masses derived under the assumption that these young massive clusters 
are virialized will overestimate the masses, in some cases to a significant extent. Should our targets be in an unresolved
cluster, the cluster mass would have been overestimated by 2.3 times. 

Note, however, that if the binarity of massive stars is independent of the cluster mass in which they formed, this effect
will be less important for very massive clusters ($>$100 OB--stars), where the cluster radial velocity dispersion will dominate over
the close binaries' velocity dispersion.

  

\subsection{Formation scenarios for massive stars}

Massive star formation still poses a significant observational and theoretical challenges, 
but evidence is mounting for the disk--accretion scenario for stars up to 30 M$_\odot$ \citep{2006astro.ph..2012B}. It is yet unclear, whether this mechanism extends to the highest stellar masses
and whether or not the stellar coalescence plays role in the formation of massive stars \citep{1998MNRAS.298...93B,2005MNRAS.362..915B}. 

When discussing the difficulty of
observationally testing these scenarios, \citet{2002hsw..work..219B}
points out the intrinsic multiplicity ratio as a potentially strong constraint. 
He notes that a high frequency of nearly-equal mass massive binaries should form in the
clusters where stellar collisions are occurring
\citep{2002hsw..work..193B}. 

Our survey demonstrates that {\em at least} 20\% and probably an even larger fraction 
of the massive stars form in compact, massive multiple systems. 
Comparison to the Monte Carlo simulations by \citet{2001RMxAC..11...29B}  adopted
to our observations, which take into account the observational biases and stellar parameter
distributions, suggest that our observations are consistent with most if not all our targets
being compact, massive binaries (Bosch, priv. comm.).

This finding may lend further support to the coalescence scenario, but may also
be consistent with the disk accretion model, whose implications on intrinsic binarity 
have not yet been explored. The significant fraction of binary stars for stellar mass ranging from
brown dwarfs \citep{2006astro.ph..2122B} through T~Tauri stars \citep{2006astro.ph..3004D} to massive stars suggests that binarity is a fundamental ingredient of star formation.

\section{Conclusions}

We present the first multi-epoch spectroscopic radial velocity survey of  
massive stars associated to young (ultra--)compact H~{\sc II} regions.
The radial velocity variations indicate massive companions in at least two of our 16
targets. The broad $\Delta$RV-distribution of our sample
resembles the distribution found by the single--epoch optical spectroscopy of the starburst
cluster R~136 \citep{2001A&A...380..137B}. The main results of
this work are:

-- We identify an O5/O6 and an O9/B1 spectral type star displaying
RV-variations as large as $\sim$90 km/s and we show that they are
close massive binary stars. 

-- Our survey demonstrates that {\em at least} 20\% of the O--stars in ultracompact~H{\sc II} regions
are in close, massive multiple systems, and may be consist with most if not all systems being compact
binaries.

-- The observed $\Delta$RV distribution of the observed sample is 35
km/s, inconsistent with single-star population. The distribution of the sub-sample
excluding the two identified massive binaries is 25 km/s, still broader than (but marginally consistent with) the distribution of single star population observed with our RV--accuracy.

-- We point out, that the presence of compact massive binaries on the overall RV--dispersion
will bias virial mass estimates of stellar clusters towards higher masses.


-- High--resolution, sensitive near-infrared spectroscopy is a powerful tool
to study the intrinsic properties of massive stars. Our RV accuracy limit is
set by the combination of the spectral resolution and the uncertainties in 
telluric line correction. Future studies capable of reaching $\sim$15 km/s or better
accuracy on a similar or larger set of targets will be able to place statistical constraints on the intrinsic frequency of compact, massive binary stars. 


\acknowledgements{The comments of the referee, Guillermo Bosch, have helped to improve the clarity of the text;
we thank him also for adopting his Monte Carlo simulations to our survey's parameter.
We thank Morten Andersen for pointing out the link to the virial mass estimates. 
We acknowledge the helpful discussions with Bringfried Stecklum and Eike G\"unther and thank Michael R. Meyer
for help with the statistical interpretation of the data set. We acknowledge the outstanding support by the Paranal staff during the observations, both in service and in visitor
mode. }
\appendix

\section{Discussion of individual objects}
\label{objects}
 
In the following we list the comments on the individual objects; a more detailed description and figures of
the individual spectra are shown in \citet{2004PhDT.........1A}.

{\bf IRAS 10049$-$5657:} Star \#324 is identified as O3V---O6.5V spectral
   type (Fig. \ref{RVSpecExam}). \hei\ absorption at 2.11 $\mu$m  is
   visible in the spectrum setting an upper limit of O5V---O6.5V to the
   spectral type. However the absence of \civ\ suggests an O3---O4V
   spectral type \citep[see][]{2005A&A...440..121B}.  The telluric line removal
   was efficient but the \brg\ line is over-subtracted due to
   overlapping nebular emission from the other dithering position. As
   its peak was unreliable, we included only the wings of the \brg\
   line in our cross-correlation. 

Star \#411 is of earlier spectral type (O3V---O4V) as no \hei\ line is
 detected and has weaker spectral lines. Although the telluric
 subtraction is in general good, the Br$\gamma$ line is contaminated
 by some telluric residuals and suffers nebular line over-subtraction.
 The \niii\ line is broad and not well--defined. However,
 cross-correlations of the wings of \brg\ and the \niii\ line gives
 very similar results to that from the correlation of only the \niii\
 line, proving the reliability of the fit.  The third epoch
 measurement gives a relative shift as large as $\sim$88 km/s.

{\bf IRAS 15408$-$5356:} Star \#~1410 is an early--type star with an
estimated spectral type of O5V---O6.5V. Although some fringing and
telluric lines are present, the cross-correlation function has a sharp
maximum. Star \#~1454 is an O8V---B2.5V spectral type star with the
He{\small I} line in absorption. The equivalent width of the \brg\ line is very
close to the division line between the O8V---B1V and B1V---B2.5V spectral
classes \citep[see][for more details]{2005A&A...440..121B}.  The telluric
correction is good, but the Br$\gamma$ line has some nebular line
over--subtraction. We used the \hei\ line as well as the wings of the
\brg\ line for the cross-correlation.

{\bf IRAS 16177$-$5018:} Star \#~271 has a spectral type of O8V---B2.5V
with weak He{\small I} and Br$\gamma$ absorption.  The wings of the
Br$\gamma$ line are useful for the cross--correlation. The 2nd epoch
observations is of somewhat worse quality than the 3rd epoch. Star
\#405 is an early-type star with O5V---O65.V classification.  The lack
of absorption lines and the weakness of the emission lines makes the
radial velocity measurement difficult.  Another very early-type star
is \#~1020, which is classified as of O5V---O6.5V spectral type. The
Br$\gamma$ line in its spectrum is filled in by nebular line emission
and was completely excluded from the fit.

{\bf IRAS 16571$-$4029:} Star \#~820 is of spectral type
O8V--B1V. Although the spectra suffer from rather stronger telluric
line residuals, the \brg\ and \hei\ line are however not affected. The
double-peaked \hei\ line provides a firm basis for the
cross-correlation.

{\bf  IRAS 17149$-$3916:} Star \#~895 has been classified as
of O5V---O6.5V spectral type.  The \civ, \niii\ and the
\brg\ line are present.  
 
{\bf IRAS 17258$-$3637:} Object \#~1558 is an O8V---B2.5V spectral type star.
 The spectra suffer from telluric residuals; in the second epoch
\brg\ is contaminated by nebular emission. Star \#~378 is classified
as O8V---B1V spectral type. Although the telluric line removal is not
good, the atmospheric lines are free from residuals. The spectra from
the two epochs are obviously shifted, while the telluric lines
coincide. This star is a strong candidate for being a binary star.

{\bf IRAS 18006$-$2422:} Star \#~770 is an O8V---B1V type star. The
spectra are of good quality apart from some telluric residuals.

{\bf IRAS 18449$-$0158:} Star \#~319 is an early type with an O5V---O6V
classification. The \brg\ line has a strange shape, might be affected
by telluric residuals. Still, the \niii\ and \hei\ line give strong
basis for the fit.  In the third epoch spectrum the \brg\ line
looks normal, but the overall telluric residuals are slightly worse.
 
{\bf IRAS 18507$+$0110:} The spectra of star \#~262 shows some fringes,
but the telluric removal was successful. The spectral classification
of this star is O5I---O8I, and it is likely not a main sequence star but
possibly a super--giant or possesses an unusually strong stellar wind. The
star \#~373 has been observed two times with two different slit
orientations the night of 19th June 2002.  This B1V---B2.5V spectral
type star provides a relatively weak cross--correlation peak due to the
improper telluric line removal. The second slit orientation on this
IRAS source included the O8V---B2.5V star \#~389.  Its spectrum shows
characteristic \hei\ absorption lines, which give a firm basis for
multi-epoch cross-correlation.

{\it Facilities:} \facility{VLT (ISAAC)}.

\bibliographystyle{aa}
\bibstyle{aa}

\bibliography{lit}
\end{document}